# Pressure induced isostructural phase transition in biskyrmion host hexagonal MnNiGa

*Anupam K. Singh[1], Parul Devi[2], Ajit K. Jena[3], Ujjawal Modanwal[1], Seung-Cheol Lee[3], Satadeep Bhattacharjee[3], Boby Joseph[4], and Sanjay Singh\*,[1]*

[1]School of Materials Science and Technology, Indian Institute of Technology (Banaras Hindu University), Varanasi-221005, India

[2]Dresden High Magnetic Field Laboratory, Helmholtz-Zentrum Dresden – Rossendorf, Bautzner Landstr. 400, 01328 Dresden, Germany

[3]Indo-Korea Science and Technology Center (IKST), Bangalore 560065, India

[4]Elettra-Sincrotrone Trieste, Strada Statale 14, Km 163.5 in Area Science Park, Basovizza 34149, Italy

Email*: ssingh.mst@iitbhu.ac.in

Magnetic skyrmions are vortex-like spin textures, which can be manipulated by external stress or pressure via magnetoelastic effects. Here, we present the observation of isostructural phase transition in a biskyrmions host hexagonal MnNiGa at pressure P~ 4 GPa using pressure-dependent synchrotron x-ray powder diffraction (XRD) data analysis. Our XRD data reveals anisotropic compression behavior with pressure with different compression rates of the *a*-axis in the basal plane and the *c*-axis in the prismatic plane. However, the hexagonal symmetry remains unchanged for pressure up to 14 GPa. Fitting of unit cell volume with pressure using a second-order Birch–Murnagan equation of state reveals that the data to fall into two distinct curves for those above and below 4 GPa. The present study contributes to the understanding of crystal structure with the application of hydrostatic pressure in the biskyrmion host MnNiGa, wherein the skyrmion textures can be manipulated by pressure due to their magnetoelastic character.

## 1. Introduction

Magnetic skyrmions are a kind of spin textures wherein particle-type swirling of spins around the unit sphere takes place in the nanometer length scale[1, 2]. They have been the subject of tremendous interest in recent years due to their developing potentiality in high-density information carriers, data processing, and memory devices[3-7] owned by their unique properties such as nanometer size, topological stability, and low power consumption (low current density ~ $10^5$–$10^6$ A-m$^{-2}$) required to drive the skyrmion textures[8-11]. Among various types of existing skyrmion textures[2], biskyrmions received a great deal of attention in recent years due to their size and topological charge tunability using the thickness, magnetic field, and electrical current[12]. Biskyrmions are a set of two individual skyrmions with opposite helicities (or vorticity)[13]. In general, biskyrmions have been observed in the centrosymmetric materials having strong uniaxial magnetic anisotropy, e.g., $La_{2-2x}Sr_{1+2x}Mn_2O_7$[13], MnNiGa[14], and MnPdGa[12].

Among different biskyrmion hosting materials, hexagonal MnNiGa turns out to be the most important candidate as it exhibits superstable biskyrmion textures in the wide temperature range (16–340 K)[14, 15] with a high value of ferromagnetic (FM) transition temperature $T_C$ ~ 350 K[14-16]. The present alloy system (MnNiGa) requires a comparatively low magnetic field (~ 0.25 T) to induce the biskyrmion textures at room temperature and can be stabilized even at zero magnetic field using field cooling protocols[14, 15]. MnNiGa exhibits a uniaxial magnetic easy axis along the *c*-direction below the FM $T_C$. Apart from $T_C$, an interesting spin reorientation transition (SRT), below which antiferromagnetic (AFM) components appear in the basal plane, has been observed in MnNiGa[16]. The magnetic properties of MnNiGa can be easily modified by compositional tuning[16]. A recent theoretical study proposed that magnetism is dominated by the Mn-Mn distance with competing FM and AFM interactions along the *c*-axis in MnNiGa[17]. In MnNiGa, a magnetically originated anomaly in the *c/a* ratio has been reported without any crystallographic symmetry change at SRT[16]. Such an anomaly in structural parameters at magnetic transition with preserved crystallographic symmetry is the indication of the presence of magnetoelastic effects[18] or isostructural phase transitions[18, 19]. The magnetoelastic effects provide the understanding of spin-lattice coupling[18]. The magnetoelastic effect has been reported in the skyrmions hosting B20 chiral magnets using Landau-Ginzburg free energy functional formulation[20, 21]. The stability of spin textures of skyrmion are very sensitive to external uniaxial stress or pressure, which can manipulate the stability of these spin textures through intrinsic spin-lattice coupling[22-24]. Therefore, manipulation of skyrmions with external parameter like pressure is of great importance to improve their practical applicability[25-27]. Although the stability of skyrmions by pressure-tuning has been investigated in the noncentrosymmetric materials for e.g., MnSi[26], FeGe[27], $Cu_2OSeO_3$[25, 28], there is almost negligible report on the centrosymmetric materials like biskyrmion host MnNiGa. This calls for a detailed study on the pressure-tuning of biskyrmions in the centrosymmetric MnNiGa.

Isostructural phase transition induced by pressure is an interesting phenomenon wherein the anomaly in structural parameters like lattice parameter, *c/a* ratio, or volume collapse appears without any change of the crystallographic symmetry[18, 29-31]. Such phase transitions always received significant attention as they are very seldom. In general, the isostructural phase transition is associated with the change in electronic structure, which reveals anomalies in the frequency of associated phonon mode[30, 31]. The electronic structure can be easily modified with pressure, which may lead to significant changes in isostructural phase transitions[30, 31].

In MnNiGa, the maximum density of biskyrmions has been reported at $T_{SRT}$ ~ 200 K[14], where AFM components increases significantly due to spin canting[16]. The application of hydrostatic pressure at room temperature can compress the lattice and result into reduced Mn-Mn interatomic distances, which may lead to the modification of magnetic interactions in MnNiGa as observed in skyrmion host $Cu_2OSeO_3$[28]. In addition, increasing behavior of FM $T_C$ with hydrostatic pressure (increasing rate of $dT_C/dP$ ~ 1.7 K/kbar)[32] has been reported for a similar hexagonal sister compound MnPtGa, which hosts skyrmions with maximum stability at its $T_{SRT}$ ~ 215 K[33]. Therefore, a natural question arises whether the SRT can be induced at room temperature driven by external pressure via modification in the magnetic interaction in MnNiGa? The answer will be helpful to stabilize the maximum density of biskyrmions at room temperature in MnNiGa and lead to low power consumption in the spintronic devices[16]. Further, the skyrmion host $Cu_2OSeO_3$ revealed dramatic enhancement in the temperature window of stable skyrmions driven by external pressure and also shows interesting pressure-driven phase transitions[28]. Therefore, a detailed pressure study of MnNiGa is required to explore the stability of biskyrmions and crystal structure under external hydrostatic pressure.

We present the observation of the pressure-driven isostructural phase transition in MnNiGa magnet using *in situ* high-pressure synchrotron XRD data analysis. The high-pressure XRD data analysis up to 14 GPa reveals the continuous reduction of lattice parameters with nonlinearity above 4 GPa. The different compression rates of the *a*-axis in the basal plane and the *c*-axis in the prismatic plane indicate the anisotropic compression behavior. However, the hexagonal symmetry remains unchanged with pressure; the fitting of lattice volume with pressure using second-order Birch–Murnagan equation of state reveals an isostructural phase transition at P ~ 4 GPa. The present results provide the understanding of the crystal structure of biskyrmion host MnNiGa with the application of hydrostatic pressure.

**2. Methods**

The polycrystalline sample of MnNiGa with nominal composition was prepared by the standard arc melting method[34]. The appropriate quantity of each constituent element with minimum 99.99% purity were melted several times to get uniform composition. After melting, weight loss was found to be below 1%. The as melted button-shaped ingot was taken in an evacuated quartz ampoule filled with argon gas and annealed at 800ºC for 6 days to get the homogeneous composition, and finally quenched in the ice water mixture[35]. To check the chemical composition, energy dispersive analysis of x-rays (EDAX) characterization was performed using EVO-Scanning Electron Microscope MA15/18 (ZEISS) equipped with an energy dispersive spectroscopy detector (Model No. 51N1000-EDS System) in the backscattered electron mode. The average composition was found to be $Mn_{1.05}Ni_{0.95}Ga$, which corresponds to MnNiGa. The temperature-dependent ac-susceptibility data at a drive field 10 Oe for 333.33 Hz frequency was collected during the warming cycle on zero field cooled sample using a superconducting quantum interference device-based magnetometer (MPMS, Quantum Design). The *in situ* high-pressure synchrotron x-ray powder diffraction (XRD) measurements were carried out using x-rays with the wavelength of 0.49584 Å at Xpress beamline at Elettra, Trieste, Italy[36]. The data was collected up to 14 GPa using membrane-driven diamond anvil cell (DAC) for generating the pressure. The methanol-ethanol mixture in 4:1 ratio was employed for the pressure transmitting medium. The pressure was monitored by the ruby fluorescence method, wherein a few tiny ruby chips (~5–10 μm) were included along with the powder sample in the DAC pressure chamber. The diffraction data were detected using PILATUS3S-6M detector. The 2D image was integrated into 1D (intensity vs 2$\theta$) diffraction data using fit2D software. The temperature dependent (300 to 15 K) laboratory source XRD patterns were collected using an 18-kW Cu rotating anode-based diffractometer attached with a closed-cycle He refrigerator. The average crystal structure was determined by the Rietveld refinements[37] using the XRD data. The refinement was carried out using FULLPROF package[38] in the *P6₃/mmc* space group (No. 194), considering all the atoms at special Wykoff positions, i.e., Mn at 2a (0, 0, 0), Ni at 2d (1/3, 2/3, 3/4) and Ga at 2c (1/3, 2/3, 1/4)[14, 16]. The variation of unit cell volume with pressure was modeled with Birch–Murnagan equation-of-state (B-M EoS)[39, 40] using the EoSFit7-GUI software[41]. The B-M EoS are given as follows:

$$P(V) = \frac{3B}{2}\left[\left(\frac{V_0}{V}\right)^{7/3} - \left(\frac{V_0}{V}\right)^{5/3}\right] \times \left\{1 + \frac{3}{4}(B' - 4)\left[\left(\frac{V_0}{V}\right)^{2/3} - 1\right]\right\}$$

In the above equation, $V$, $V_0$, $B$, and $B'$ are volume, reference volume, bulk modulus, and pressure derivative of bulk modulus, respectively[42]. The second-order B-M EoS can be obtained by considering $B' = 4$ in the above equation[43, 44].

## 3. Results and Discussion

### 3.1. Phase purity and phase transition

The Rietveld refinement using the XRD data at ambient condition is shown in **Figure 1**(a), which shows an excellent fit between observed and calculated profiles by accounting for all the Bragg peaks. This confirms that the MnNiGa crystallizes into the hexagonal structure with $P6_3/mmc$ space group. The refined lattice parameters were found to be $a = b = 4.1574(1)$ Å, $c = 5.3261(1)$ Å. The present diffraction pattern and refined lattice parameters are in good agreement with the previous report[14, 16]. Thus, XRD data at ambient conditions confirm the phase purity of the MnNiGa sample.

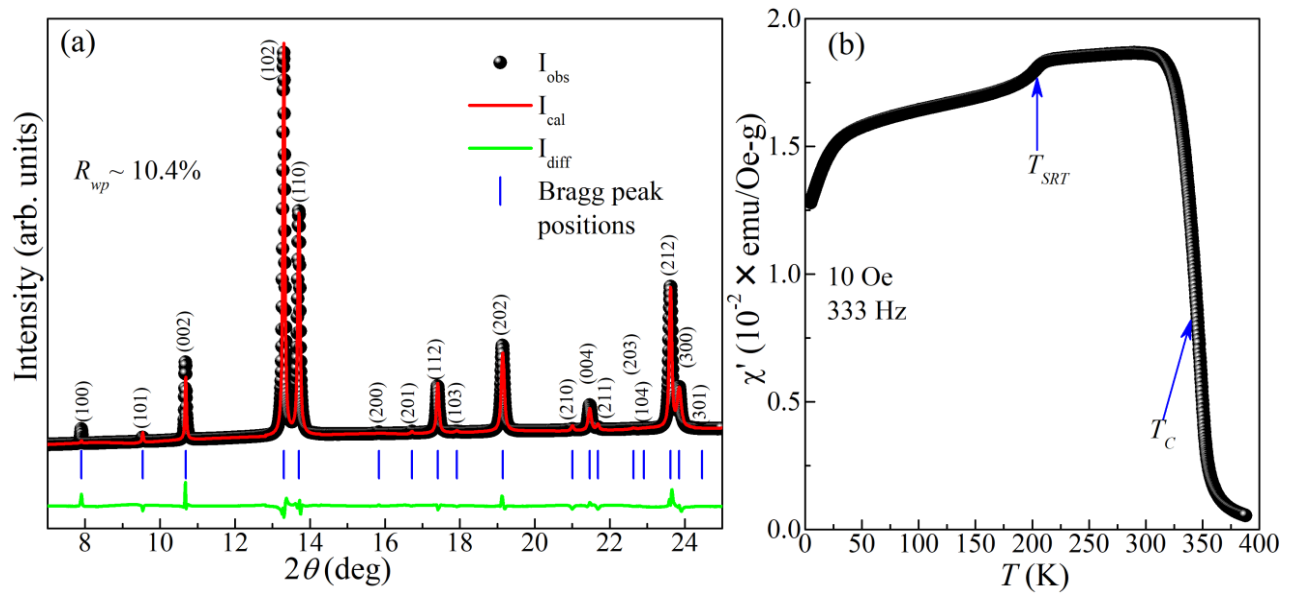

**Figure 1.** (a) Result of Rietveld refinement using synchrotron x-ray powder diffraction pattern at the ambient condition of MnNiGa. The "$R_{wp}$" represents the weighted agreement factor of refinement. The miller indices are given above each reflection. (b) Temperature dependence of real part of ac-susceptibility ($\chi'$) of MnNiGa. The "$T_C$" and "$T_{SRT}$" indicate ferromagnetic and spin reorientation transition temperatures, respectively.

The temperature-dependence of the real part of ac-susceptibility ($\chi'(T)$) is shown in **Figure 1**(b). On lowering the temperature, the sudden increase in $\chi'(T)$ around 350 K corresponds to the paramagnetic phase to ferromagnetic (FM) phase transition. The value of FM $T_C \sim 347$ K was determined by taking minima of first-order differentiation of $\chi'(T)$. Besides FM $T_C$, the $\chi'(T)$ decreases gradually below 200 K (see **Figure 1**(b)) due to the presence of spin reorientation transition (SRT)[16]. The SRT temperature was found at $T_{SRT} \sim 200$ K as calculated by the temperature derivative of $\chi'(T)$. However, a rapid decrease in $\chi'(T)$ can be noted below 50 K, which is a matter of future investigation. The present behavior of $\chi'(T)$ and magnetic transition temperatures are in good agreement with the previous report[16].

## 3.2. Structural investigation under hydrostatic pressure

The XRD pattern at different pressure is shown in **Figure 2**(a), whose enlarged view around the most intense Bragg peak region is depicted in **Figure 2**(b). In **Figure 2**, the absence of any splitting or appearance of additional Bragg peaks with pressure manifests that the hexagonal symmetry remains preserved. This indicates the absence of any structure phase transition in MnNiGa with pressure up to 14 GPa, the maximum pressure up to which data has been collected.

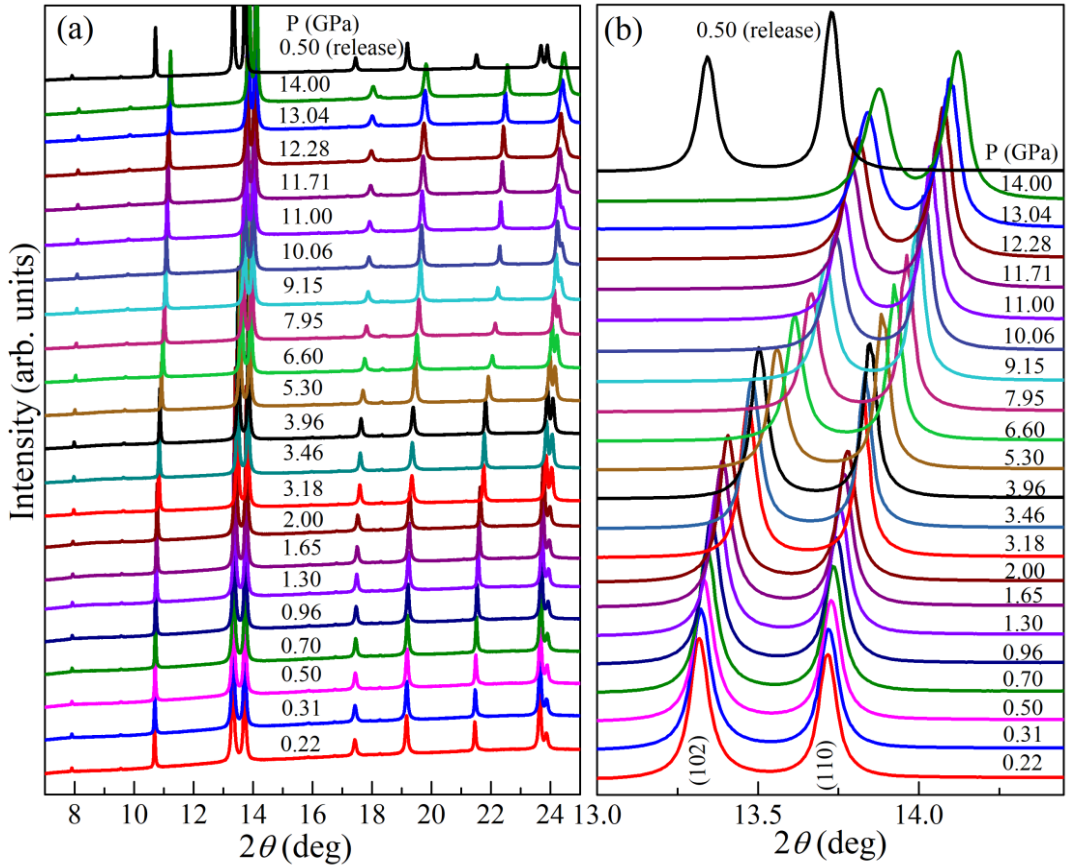

**Figure 2.** (a) The synchrotron x-ray powder diffraction patterns collected at various pressure (indicated) up to 14 GPa of MnNiGa. (b) The enlarged view of (a) around the most intense

Bragg peak region. The miller indices of both major peaks are given in the bottom-most in (b). The "P (GPa)" represents the pressure in the GPa unit. The top most pattern labelled by "0.50 (release)", was collected during releasing the pressure with effective pressure of 0.50 GPa.

Further, the systematic shifting of the Bragg peaks to the higher 2$\theta$ side with pressure suggests the compression of lattice parameters or interplanar spacings (see **Figure 2**). It is important to mention here that, however, the intensity of (110) reflection should be higher than (102) as observed in the computed diffraction pattern, the reverse situation at the ambient condition (see **Figure 1**(a)) as well as at lower pressure (see **Figure 2**(b)) is certainly due to texturing effect. At P ≥ 3.96 GPa, the intensity of (110) reflection became higher than (102) reflection (intensity switching), which suggests the texturing effect is suppressed at higher pressure (see **Figure 2**(b)). All these behaviors in the data are found to be reversible with pressure, as demonstrated by a comparison of the bottom XRD pattern, which was collected at 0.22 GPa during pressure application, and the top XRD pattern, which was collected at 0.5 GPa during pressure release. This indicates reversibility of structure with the pressure effects, i.e., after releasing the pressure, complete recovery of the structure to the ambient conditions. This reversibility behavior with pressure in MnNiGa is in marked contrast to other skyrmion host $Cu_2OSeO_3$, wherein the irreversibility behavior with pressure has been observed[28]. The reversibility behavior of MnNiGa makes this alloy system ideal for pressure-tuning of skyrmion textures.

The lattice parameters are obtained from the Rietveld refinement using high-pressure XRD data. The evolution of lattice parameters (*a* and *c*) with pressure is shown in **Figures 3**(a) and (b). However, the behavior of the lattice parameter at low pressure shows linear compression as indicated by a linear fit to the observed lattice parameter, a clear departure from the linearity observed above 4 GPa in the form of deviation between extrapolated linear fit and the data in **Figures 3**(a) and (b). The slope of linear fit to in-plane (*a*) and out-of-plane (*c*) lattice parameter was found to be d*a*/dP ≈ −0.01126 Å GPa$^{-1}$ and d*c*/dP ≈ −0.02306 Å GPa$^{-1}$, respectively (see **Figures 3**(a) and (b)). This means a higher compression rate for the out-of-plane *c*-axis in comparison to the in-plane *a*-axis with pressure, i.e., the presence of anisotropic compression behavior in the MnNiGa as reported in the other hexagonal system[30].

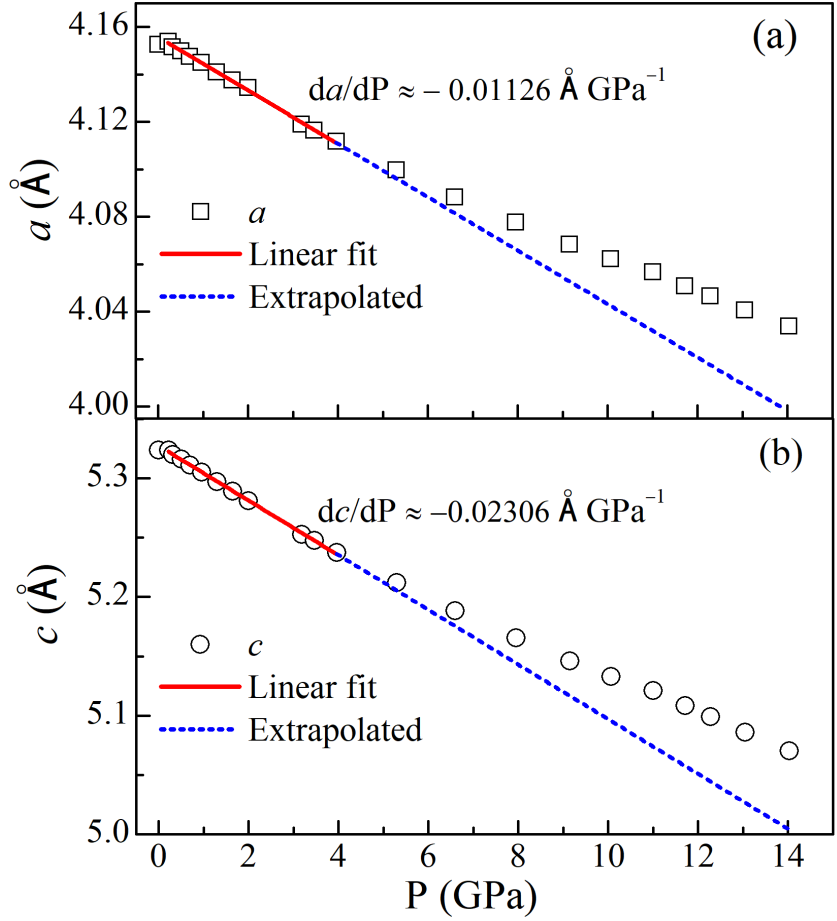

**Figure 3.** Pressure-dependence of (a) in-plane and (b) out-of-plane lattice parameters of MnNiGa. The red line and dotted blue line in (a) and (b) represent the linear compression and extrapolated region of linear compression behavior, respectively. The "d$a$/dP" and "d$c$/dP" indicate the linear compression rate in the $a$ and $c$-parameter in (a) and (b), respectively. The error in lattice parameters is smaller than the symbol size.

In addition to the lattice parameters, the deviation from the systematic linearity with pressure above 4 GPa was also observed in the unit cell volume ($V$) and $c/a$ ratio as depicted in **Figure 4** and in the inset of **Figure 4**, respectively. It is important to mention here that we have a large no. of data points at both low and high pressure regions, such kind of detailed analysis (i.e., reliable slope change in the $c/a$ ratio) is possible. The decreasing behavior of the $c/a$ ratio with pressure further manifests that the $c$-axis is more compressible than the $a$-axis (see the inset of **Figure 4**). This behavior is usually expected for anisotropic crystals due to the weak van der Waals interlayer forces along the $c$-direction[30].

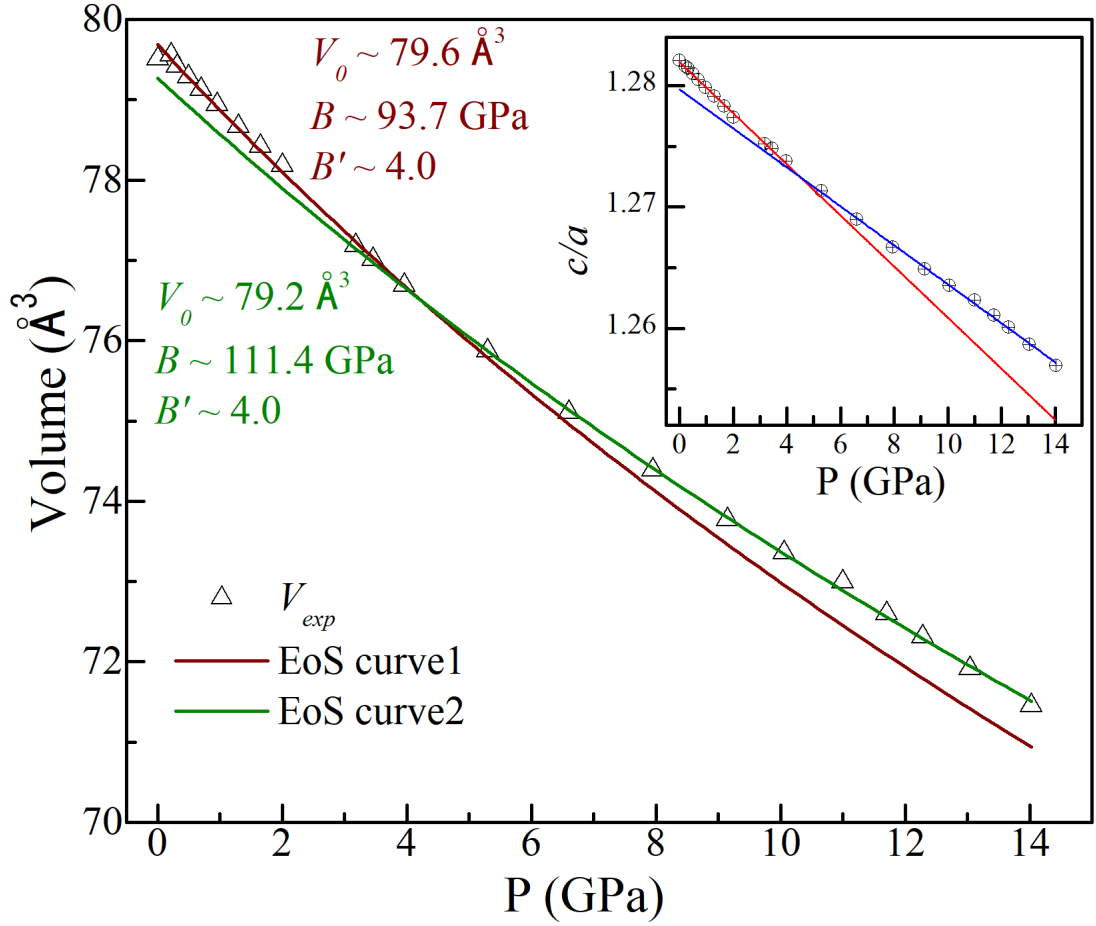

**Figure 4.** Pressure-dependence of unit cell volume of MnNiGa. Solid lines indicate the results of a second-order Birch–Murnagan equation-of-state (EoS) fit to the data. "EoS1" and "EoS2" represent the fitting considering the data upto 4 GPa and above 4 GPa, respectively. The $V_0$, $B$ and $B'$ are the parameters obtained from the fit. Pressure-dependence of the $c/a$ ratio is depicted in the inset, wherein red and straight blue lines indicate the plot from lower to higher and higher to lower pressure region, respectively.

The significant deviation in the structural parameters ($a$, $c$, $c/a$, and $V$) above 4 GPa from the linearity without any change in symmetry is the indication of isostructural phase transition[30, 31, 42, 45-48]. In order to investigate it, a second-order Birch–Murnagan equation-of-state[39, 40] (2nd O B-M EoS) was employed to model the evolution of volume with pressure (P-V). Since considering the whole P-V region together did not yield a satisfactory fit, two separate regions of the P-V plot were considered for the fitting using two independent 2nd O B-M EoS[49]. The first region was considered at the lower-pressure side (up to 4 GPa), while the second was at the higher-pressure side (above 4 GPa). The result of the fitting is shown in **Figure 4**, which yield different bulk modulus ($B$) for different region. The value of $B$ was found to be increased from ~93.7 GPa to ~111.4 GPa from low to high-pressure regions. In skyrmion host $Cu_2OSeO_3$, cubic to monoclinic phase transition has been observed at 7 GPa with $B$ increased from 74.8 to 161.1 GPa[28]. Compared to this, a relatively smaller change in $B$ (~93.7 to ~111.4 GPa) of MnNiGa manifests the pressure-induced isostructural phase transition above 4 GPa and suggests that the high-pressure phase has lesser compressibility than low-pressure phase in MnNiGa as reported in other systems (like PdPS[42]). The anomaly above 4 GPa is also observed in the linearization of BM-EOS with the Eulerian strain[50,51] (see Figure S1 and related texts on page 1 of the Supporting Information). Interestingly, our theoretical results (electronic band structure, density of states at the Fermi level and bulk modulus) are found to be in excellent agreement with the experimental findings (see Figures S2, S3 and related texts on pages 2 to 3 of the Supporting Information). The presence of pressure-induced isostructural phase transition is of fundamental importance and opens a new pathway for future investigation of pressure-tuning of skyrmion textures in the centrosymmetric MnNiGa and related system.

The maximum density of biskyrmions has been reported at $T_{SRT}$ ~ 200 K in hexagonal MnNiGa[12]. We compared the lattice parameters (LP) obtained from pressure and temperature dependent XRD (see Figures S4, S5 and Table S1 of the Supporting Information), which reveals that LP at 200 K (~$T_{SRT}$) are close to the LP at 0.7 GPa. This indicates that $T_{SRT}$ or maximum density of biskyrmions can be stabilized at room temperature by using P ~ 0.7 GPa in hexagonal MnNiGa.

## 4. Conclusion

In summary, we have shown pressure-induced isostructural phase transition in the hexagonal MnNiGa using high-pressure synchrotron XRD analysis. The XRD pattern at ambient condition and ac-susceptibility measurements confirm the phase purity and magnetic phase transitions, respectively. The *in situ* high-pressure XRD data analysis up to 14 GPa reveals the continuous reduction of lattice parameters with nonlinearity above 4 GPa. The different compression rates of the *a*-parameter (in-plane) and *c*-parameter (out-of-plane) indicate the anisotropic compression behavior. However, the hexagonal symmetry remains unchanged with pressure; the fitting of lattice volume with pressure using second-order Birch–Murnagan equation of state reveals an isostructural phase transition around 4 GPa. A comparison of pressure and temperature dependent lattice parameters suggests that SRT can be induced by ~0.7 GPa pressure at room temperature. The present results open a new window for future investigation of pressure-tuning of skyrmions in the centrosymmetric MnNiGa and related systems.

**Acknowledgments**

S.S. is thankful to the Science and Engineering Research Board of India for financial support through the award of Ramanujan Fellowship (Grant No: SB/S2/RJN-015/2017). S.S. thanks UGC-DAE CSR, Indore, for financial support through the "CRS" Scheme. The major portions of this research were conducted using the light source of Elettra-Sincrotrone Trieste. BJ acknowledges the Xpress plus internal project of Elettra-Sincrotone Trieste. DST, Govt. of India, is acknowledged for support in performing the high-pressure diffraction measurements at the Xpress beamline of the Elettra (proposal no. 20205391).

# Supporting Information

**Pressure induced isostructural phase transition in biskyrmion host hexagonal MnNiGa**

*Anupam K. Singh[1], Parul Devi[2], Ajit K. Jena[3], Ujjawal Modanwal[1], Seung-Cheol Lee[3], Satadeep Bhattacharjee[3], Boby Joseph[4], and Sanjay Singh\*[1]*

**Linearization of Birch–Murnagan equation-of-state (BM-EOS) vs the Eulerian strain:**

In order to analyze the experimental observations more accurately, we carried out the linearization of Birch–Murnagan equation-of-state (BM-EOS) vs the Eulerian strain ($f_E$)[1, 2]:

$$H = B + \frac{3B}{2}(B' - 4)f_E,$$

Where *H* is the reduced pressure with $H = \frac{3P}{3 f_E (1+2 f_E)^{5/2}}$ and $f_E = \frac{1}{2}\left[\left(\frac{V_0}{V}\right)^2 - 1\right]$. In the above equations, P, V, $V_0$, B, and B' are pressure, volume, reference volume, bulk modulus, and pressure derivative of bulk modulus, respectively[1, 2]. The *H* vs $f_E$ plot should be linear in the absence any phase transition[1, 2]. However, the *H* vs $f_E$ plot for the hexagonal MnNiGa is shown in **Figure S1**, which clearly reveals a curvature change in the pressure range of ~2 to 3.96 GPa. After 3.96 GPa, it becomes almost linear (see **Figure S1**). The change in the *H* vs $f_E$ plot further confirms the presence of isostructural phase transition above 4 GPa in the present hexagonal MnNiGa as reported in the other system on the basis of *H* vs $f_E$ plot[2].

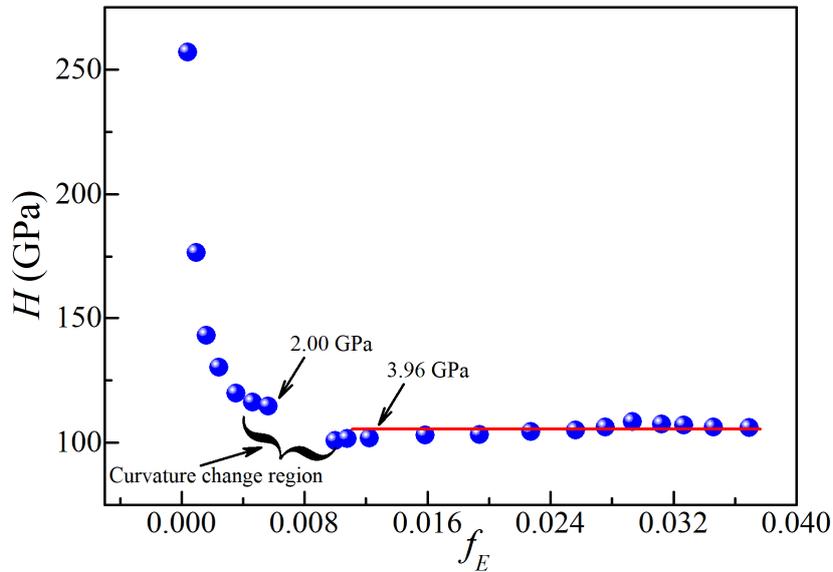

**Figure S1**: The reduced pressure (*H*) vs Eulerian strain ($f_E$) plot. The red line shows linear behavior from higher to lower pressure regions. The curvature change region is indicated by a curly bracket. Arrows indicate the 2.00 and 3.96 GPa pressure points. The error in the *H* is smaller than the symbol size.

**Theoretical Calculations:**

In order to further confirm the experimental observations of isostructural phase transition induced by pressure in the hexagonal MnNiGa, we performed the density functional theoretical calculations to obtain the electronic band structure, density of states (DOS) and bulk modulus (*B*) as a function of pressure. We have employed pseudo-potential based density-functional theory as implemented in Quantum ESPRESSO (QE)[3] within the framework of generalized gradient approximation (PBE-GGA)[4] to solve the Kohn-Sham Hamiltonian. Optimized norm-conserved Vanderbilt's pseudopotentials[5] are used in the calculations and the kinetic energy cutoff for the planewave is taken as 80 Ry. The electronic integration over the Brillouin zone is approximated by the Gaussian smearing of 0.01 Ry for the self-consistent (SC) calculations. The Monkhorst-Pack k-grid of $8 \times 8 \times 6$ and $10 \times 10 \times 8$ are respectively used for the SC and DOS calculations. Tetrahedral occupation is considered for the DOS calculations. A tight energy threshold of $10^{-8}$ Ry has been set for the SC total energy calculations. Murnaghan equation of state, implemented within QE, has been used to calculate the bulk modulus. For different pressures, the corresponding experimental structures have been used in the theoretical calculations.

Electronic band structures of the hexagonal MnNiGa upto 10.06 GPa are shown in **Figure S2(a)** and **Figure S2(b)** for the spin-up and spin-down consideration, respectively. Although spin-down consideration shows a continuous monotonic change with pressure (see **Figure S2(b)**), there are several observations can be drawn from the spin-up case. The degenerate energy states observed for ambient, 1.3 GPa and 3.18 GPa below the Fermi level at the M-point (see the bands at 1.3 GPa and 3.18 GPa in the red-colored bounded region in **Figure S2(a)**). Interestingly, above 3.18 GPa, non-degenerate states with different energy arise (see the bands at 5.3 GPa, 6.3 GPa and 10.06 GPa below the Fermi level at the M-point in the red-colored bounded region in **Figure S2(a)**). Thus, these results indicate that the electronic structure changes significantly above the 3.18 GPa, which is a possible factor in driving the pressure-induced isostructural phase transition as observed experimentally in the present alloy system (MnNiGa).

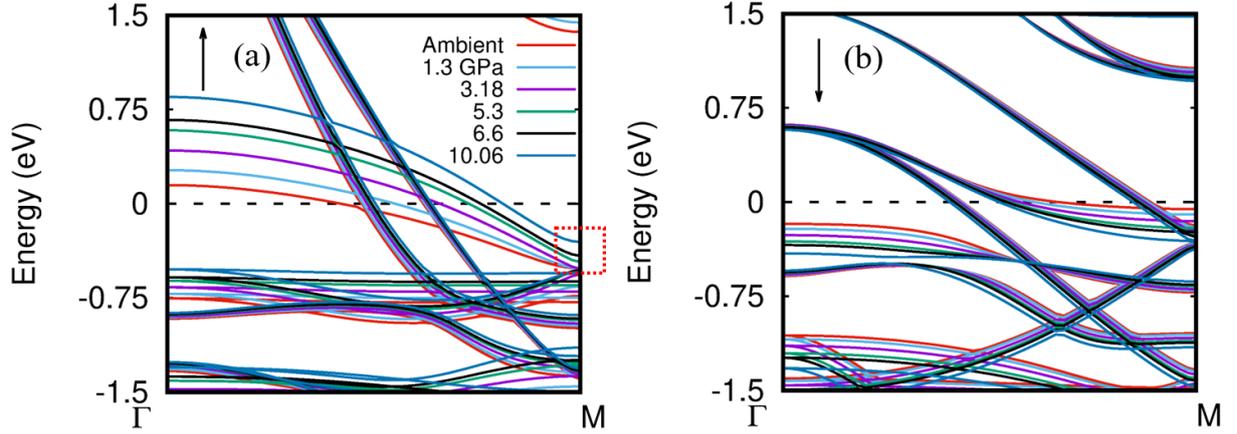

**Figure S2**: Electronic band structure of the hexagonal MnNiGa from ambient to 10.06 GPa considering the (a) up-spin and (b) down-spin. The degenerate and non-degenerate states at the M-point are guided to the eye by the red-colored bounded region in (a). The dotted horizontal line in (a) and (b) represents the Fermi level.

Theoretically obtained DOS at the Fermi level and *B* of the hexagonal MnNiGa as a function of pressure are shown in **Figure S3(a)** and **Figure S3(b)**, respectively. It is evident from **Figure S3(a)** that different decreasing behavior appears in the DOS at the Fermi level above and below ~4 GPa, as indicated by the shaded region in **Figure S3(a)**. Apart from DOS, a clear deviation from the linearity above ~4 GPa appears also in the *B* (see **Figure S3(b)**). The pressure at which the crossover occurs in the density functional results could be slightly different from the experimental ones. Therefore, all these results support the experimental finding of the presence of isostructural phase transition above 4 GPa in the hexagonal MnNiGa.

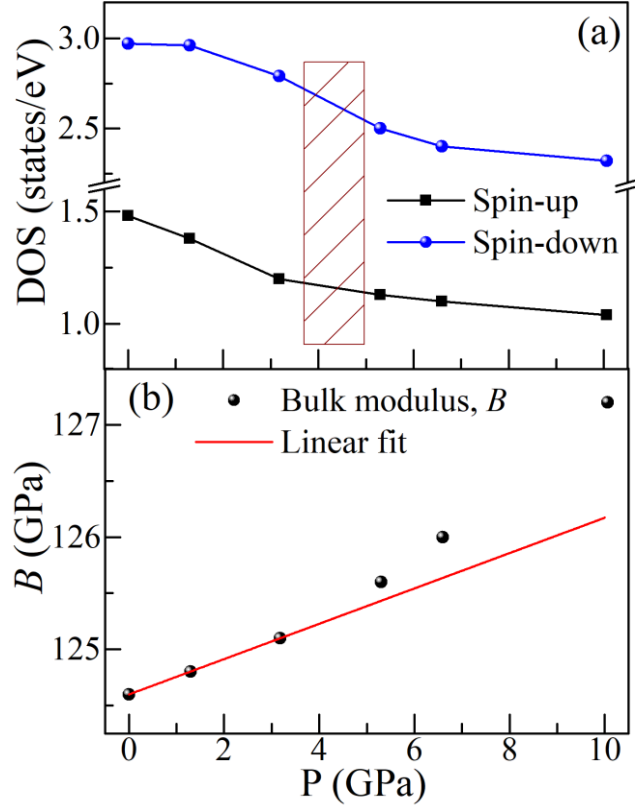

**Figure S3**: Theoretically calculated (a) density of states (DOS) at the Fermi level and (b) bulk modulus ($B$) of the hexagonal MnNiGa as a function of pressure. The shaded region in (a) is a guide for the eye to visualize the change region. The red lines in (b) indicate the linear fit.

**Temperature dependent x-ray diffraction patterns:**

The temperature dependent x-ray diffraction (XRD) patterns of MnNiGa were collected using an 18-kW Cu rotating anode-based diffractometer attached with a closed-cycle He refrigerator. The laboratory source XRD patterns in the temperature range of 300 to 15 K are shown in **Figure S4(a)**, whose enlarged view around the most intense Bragg peak region is depicted in **Figure S4(b)**. It is evident from **Figures S4(a)** and **S4(b)** that no extra peaks or any peak splitting appear down to 15 K.

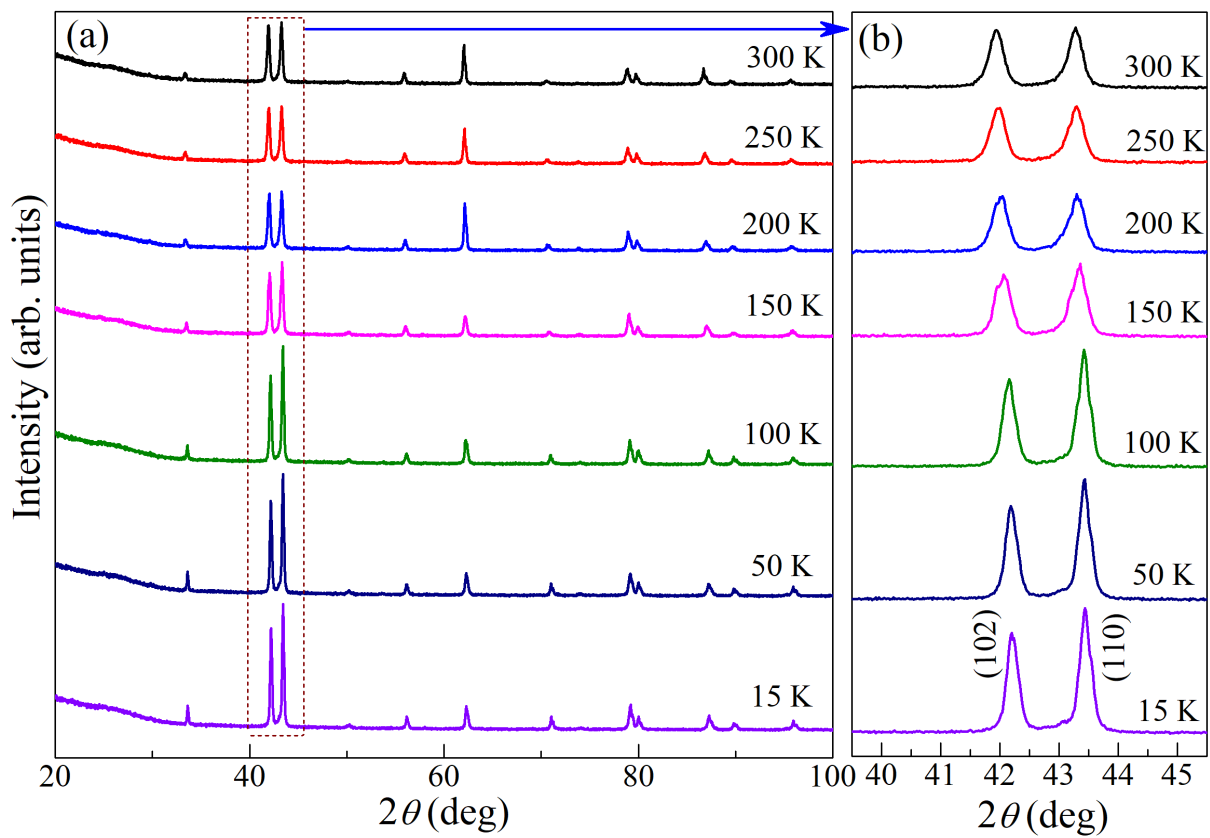

**Figure S4**: (a) The temperature dependent (300 to 15 K) laboratory source x-ray diffraction patterns of MnNiGa. (b) An enlarged view of (a) around the most intense Bragg peak region. The miller indices of both major peaks are indicated near the bottom-most pattern in (b).

The results of the Le Bail refinements using the XRD patterns at the selected temperatures with hexagonal structure in the $P6_3/mmc$ space group are shown in **Figures S5(a)** to **S5(d)**, which suggest that the hexagonal crystal structure of MnNiGa remain unchanged down to 15 K from 300 K.

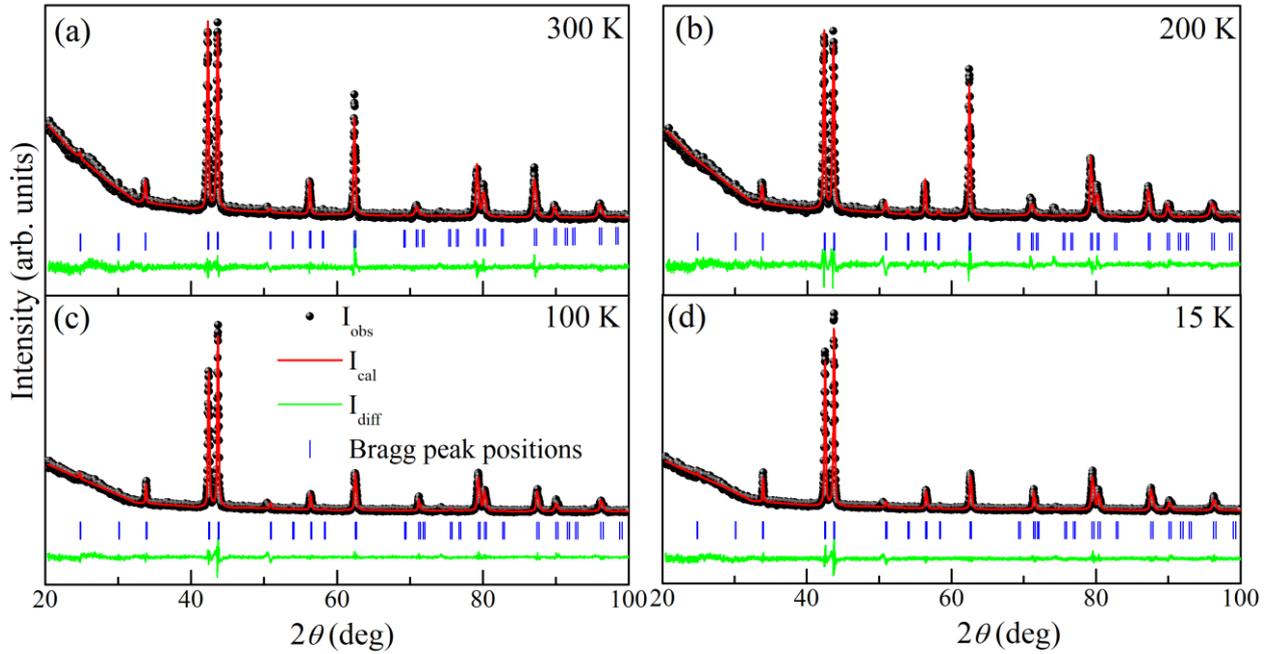

**Figure S5**: The observed profile (black spheres), calculated profile (continuous red line), difference profile (continuous green line), and Bragg peak positions (blue ticks), obtained after the Le Bail refinements with the $P6_3/mmc$ space group using laboratory source x-ray powder diffraction patterns of MnNiGa at (a) 300 K, (b) 200 K, (c) 100 K, and (d) 15 K.

It is interesting to compare the temperature and pressure dependent lattice parameters (LP) to explore the possibility of the spin reorientation transition (SRT) stabilization at room temperature induced by pressure in the hexagonal MnNiGa. A comparison of LP with temperature and pressure is given in **Table S1**, which reveals that the LP at 200 K (~$T_{SRT}$; see **Figure 1(b)** of the manuscript) almost matches with LP at P ~ 0.7 GPa pressure. This indicates that the $T_{SRT}$ can be stabilized at room temperature by applying 0.7 GPa pressure in the hexagonal NiMnGa.

**Table S1**: Comparison of temperature and pressure dependent lattice parameters (*a* and *c*) of hexagonal MnNiGa. The "$T_{SRT}$" indicates the spin reorientation transition temperature.

| Temperature | Lattice Parameters (Å) | Pressure | Lattice Parameters (Å) |
|---|---|---|---|
| 300 K | *a* = 4.1553(3) <br> *c* = 5.3245(5) | 0 GPa | *a* = 4.1574(1) <br> *c* = 5.3261(1) |
| 200 K (~$T_{SRT}$) | *a* = 4.1495(1) <br> *c* = 5.3110 (4) | 0.7 GPa | *a* = 4.1485(2) <br> *c* = 5.3118(3) |
| 100 K | *a* = 4.1459(1) <br> *c* = 5.2977(2) | 1.3 GPa | *a* = 4.1430(2) <br> *c* = 5.2990(3) |
| 15 K | *a* = 4.1428(1) <br> *c* = 5.2876(2) | 2 GPa | *a* = 4.1356(2) <br> *c* = 5.2823(3) |